\documentclass[a4paper,12pt]{article}
\usepackage[utf8]{inputenc}
\usepackage[a4paper,width=150mm,top=25mm,bottom=25mm]{geometry}
\usepackage{authblk}

\newcommand{\argmax}{\mathop{\mathrm{argmax}}}

\title{Exploring the heterogeneous impacts of Indonesia's conditional cash transfer scheme (PKH) on maternal health care utilisation using instrumental causal forests}

\author[1]{Vishalie Shah}
\author[2]{Julia Hatamyar\thanks{Corresponding author: Julia Hatamyar, email: {\tt julia.hatamyar@york.ac.uk}. Funding for this work was provided by the NIHR. Award ID: NIHR133252. Thanks to participants at COMPIE 2024 for helpful feedback.}}
\author[3]{Taufik Hidayat}
\author[4]{Noemi Kreif}

\affil[1]{\small{IQVIA UK}}
\affil[2]{Centre for Health Economics, University of York}
\affil[3]{University of Sussex \& Center for Health Economics and Policy Studies (CHEPS), Faculty of Public Health, Universitas Indonesia}
\affil[4]{University of Washington}
\date{\today}

\usepackage[sort,authoryear]{natbib}
\usepackage{fancyhdr}
\usepackage{amsmath}

\usepackage{pdflscape}
\usepackage{booktabs}
\usepackage{setspace}
\usepackage{longtable}
\usepackage[table]{xcolor} 
\usepackage{graphicx}
\usepackage{amssymb}
\usepackage{lscape}
\usepackage{longtable}
\usepackage{array}
\usepackage{makecell}
\newcolumntype{P}[1]{>{\centering\arraybackslash}p{#1}}
\usepackage{caption}
\usepackage{subcaption}
\usepackage{threeparttable}
\usepackage[shortlabels]{enumitem}
\usepackage{float}
\usepackage[title]{appendix}
\usepackage{etoolbox}
\usepackage[multiple]{footmisc}
\usepackage{adjustbox}
\usepackage{url}

\usepackage{tikz}
\usetikzlibrary{arrows.meta,bending,positioning}

\usepackage{lipsum}
\begin{document}
\maketitle
\singlespacing
\begin{abstract}
This paper uses instrumental causal forests, a novel machine learning method, to explore the treatment effect heterogeneity of Indonesia's conditional cash transfer scheme on maternal health care utilisation. Using randomised programme assignment as an instrument for enrollment in the scheme, we estimate conditional local average treatment effects for four key outcomes: good assisted delivery, delivery in a health care facility, pre-natal visits, and post-natal visits. We find significant treatment effect heterogeneity by supply-side characteristics, even though supply-side readiness was taken into account during programme development. Mothers in areas with more doctors, nurses, and delivery assistants were more likely to benefit from the programme, in terms of increased rates of good assisted delivery outcome. We also find large differences in benefits according to indicators of household poverty and survey wave, reflecting the possible impact of changes in programme design in its later years. The impact on post-natal visits in 2013 displayed the largest heterogeneity among all outcomes, with some women \textit{less} likely to attend post-natal check ups after receiving the cash transfer in the long term. 
\end{abstract}
\hspace{10pt}

\doublespacing

\clearpage
\section{Introduction}
In recent decades, conditional cash transfer (CCT) programmes have become a popular policy tool in many low- and middle-income countries for alleviating short-term poverty via cash injections, while also improving the long-term trajectory of vulnerable families via investments in human capital \citep{Parker2017-kv}. Regular cash payments are made to households in exchange for compliance with certain behaviours, such as school attendance for children, or attendance at health check-ups for new mothers, among others. Numerous evaluations of CCTs, mainly based on randomised experiments (for example, \textit{PROGRESA} in Mexico and \textit{PRAF} in Honduras), have demonstrated the ability of these interventions to make substantial improvements on education, consumption and health outcomes, particularly in the short-term \citep{Fiszbein2009-gl,Millan2019-ps,Bastagli2019-jt,Owusu-Addo2018-yo,Garcia2017-ac,Lagarde2007-hb,Kabeer2015-jt}.

The majority of the CCT evaluation literature to date has focused on average effects, with fewer studies formally analysing whether effects differ for population subgroups defined by observable characteristics. Anti-poverty programmes are expected to impact households differently depending on their ability to convert the cash injections into desirable outcomes, which is highly dependent on their own attributes \citep{Ravallion2005-yx,Cooper2020-bx}. For example, urban mothers may already have easier access to preventive health care facilities to satisfy the health requirements for pre- and post-natal check-ups, compared to those in rural regions. The cash injection could assist rural households in addressing some of the financial barriers in accessing health care, such as transport costs. Understanding this type of heterogeneity in programme impacts can help to inform better policy targeting that, among other objectives, identifies households that are expected to benefit the most, and protects those that are expected to benefit the least \citep{Cooper2020-bx}. Of those studies that do in fact explore subgroup effects for pre-specified populations, many find evidence of heterogeneity that is consistent with the broader literature suggesting that CCT effectiveness on health outcomes is modified through various social determinants of health, such as education, wealth and the urban-rural distinction \citep{Owusu-Addo2018-yo,Bastagli2019-jt}. 

In this paper, we contribute to the growing evidence base on the heterogeneous effects of CCT programmes by evaluating Indonesia's \textit{Program Keluarga Harapan} (Family Hope Programme, or PKH) using a unique data set from a large-scale randomised experiment that was implemented in 2007 alongside a baseline survey and two follow-up surveys in 2009 and 2013. We are interested in exploring how enrolment into PKH  has influenced maternal health care utilisation in the short-term (2009) and the longer-term (2013) by performing separate analyses for both time periods. Existing evaluations of PKH have focused on estimating overall average effects, finding notable improvements in various utilisation outcomes, such as the probability of having a facility delivery, or that the delivery is assisted by trained professionals \citep{Kusuma2016-tx,Cahyadi2020-lw}. Few studies have acknowledged that PKH impacts may be heterogeneous, and preliminary subgroup analyses that stratify treatment effects by pre-selected ``effect modifiers" (e.g. gender, employment sector, parental education levels), have shown this to be the case. However, traditional approaches to heterogeneous treatment effect estimation (e.g. estimating treatment effects on effect modifier strata, or performing interactions between the treatment variable with effect modifiers in a linear regression) have their own limitations, including potentially arbitrary subgroup analyses and issues of multiple hypotheses testing. Recent developments in machine learning (ML) based estimators of treatment effect heterogeneity use flexible modelling strategies that can identify heterogeneous population subgroups in a more principled way by performing (higher-order) interactions between the treatment variable and baseline characteristics. Generalised random forests, developed by \citet{Athey2019-qf}, have become a popular tree-based ML tool for estimating causal effects, including the conditional average treatment effect (CATE) function, which captures heterogeneity in treatment effects, by searching over the entire covariate space in a data-adaptive manner, rather than focusing on specific covariates selected \textit{a priori}.

We rely on the random assignment of PKH to inform our empirical strategy. While the programme was randomised, actual enrolment was not random, due to non-compliance and targeted assignment within randomised populations. Following \citep{Cahyadi2020-lw}, we address these potential observed and unobserved differences between enrolled and not enrolled groups using an instrumental variable analysis, where we instrument PKH enrolment with theoriginal randomisation mechanism itself. We extend their analysis by focussing on estimating the heterogenous effects of the programme, using the instrumental causal forest approach \citet{Athey2019-qf}; a variant of generalised random forests that allows for the presence of unmeasured confounding if there is a valid instrument available. This method  targets the estimation of the so-called conditional local average treatment effect, characterising how treatment effects vary according to observed characteristics of compliers, in our case mothers who complied with the randomisation protocol. We summarise treatment effect heterogeneity using three approaches: (1) we find the best linear predictors of heterogeneity; (2) we assess how the most and least affected population groups differ in terms of observable characteristics, and (3) we estimate depth-two optimal policy trees and describe which characteristics are chosen as the most important decision criteria for treatment allocation
\citep{Chernozhukov2018-ef,Semenova2021-lp,Knaus2021-zb,Kennedy2020-ym,Athey2019-mk}.  

This paper has three main contributions. First, we add to the growing collection of CCT evaluation studies that look beyond average impacts and capture heterogenous impacts according to observable differences in covariates. A novel contribution is our use of data-driven methods, in particular tree-based causal ML, to estimate and make inferences on the heterogeneous impacts of a CCT intervention. Our findings could help to support those from existing heterogeneity analyses by identifying potentially new population subgroups that have not been specified in advance. Second, to our knowledge, this is the first paper to evaluate a large-scale policy intervention using instrumental forests. Several published papers have used causal forests\footnote{Causal forests are a variant of the generalised random forests framework that estimate the CATE function.} without incorporating an instrumental variable analysis to address endogeneity concerns \citep{Kreif2022-uj,Bertrand2017-ol,Davis2017-zo,ONeill2018-lg,Hoffman2019-bg,Athey2019-mk}.

\section{The PKH programme}
\subsection{Background and design}
PKH was launched by the Government of Indonesia in 2007 as the country's first CCT programme targeted to households. It was designed in response to increasing concerns around the country's consistently poor human development outcomes (i.e. high mortality rates for new mothers and children under-5 and low enrolment rates for primary and secondary schools) compared to neighbouring countries, despite experiencing sustained economic growth. Prior to the implementation of PKH, an unconditional cash transfer programme (\textit{Bantuan Langsung Tunai}, or BLT) was trialed but failed to achieve the desired outcomes due to ineffective targeting of the poor and a lack of conditions on the transfers to incentivise poverty-reducing behaviours \citep{World_Bank2012-dd}. In comparison, PKH provides quarterly cash transfers to extremely poor households with pregnant women and/or children, with the objective of improving lagging health and education outcomes \citep{Alatas2011-di}. The cash payments, ranging between 600,000 and 2,200,000 rupiah per quarter (approximately 60 to 330 US dollars, depending on household composition) were made to women in the household, who were informed at the start of the programme that in order to continue receiving payments, they must fulfil certain obligations. For example, pregnant or lactating women are required to make four pre-natal care visits and two post-natal visits, take iron tablets during pregnancy, and have an assisted delivery with a trained professional.\footnote{The programme also required that children receive immunizations and attend school.} The average duration of household enrolment into PKH is between two to four years, in which time the programme aims to achieve improvements in welfare and human development indicators. 

In the first phase of the experimental rollout, PKH was introduced in six provinces -West Java, East Java, North Sulawesi, Gorontalo, East Nusa  Tenggara, and  the capital city of Jakarta - depending on their willingness to participate and how well they represent Indonesia's diversity in terms of levels of deprivation, the urban-rural distinction and remoteness. The richest 20 percent of districts within each province were excluded, and among the remaining districts, 736 subdistricts (corresponding to a population size of 36 million) were randomly selected to participate in the trial based on their supply-side readiness to deliver adequate health and education services. Of these participating subdistricts, 438 were randomly selected for the treatment group, and further to this, 700,000 households classified as being extremely poor were targeted for enrolment into PKH. The selection process involved applying a proxy means test to all poor households to identify those poor enough to be included in the beneficiary list.

The World Bank collected data via a baseline survey in the months prior to launch, and two follow-up surveys in 2009 and 2013. Out of the 736 sampled subdistricts, 360 were randomly chosen for data collection (corresponding to approximately 14,000 households), which included beneficiary and non-beneficiary households in 180 treated subdistricts, and eligible households in 180 control subdistricts. The sampling frame was created by randomly selecting villages within each subdistrict, and then selecting one subvillage within each village. Four households were randomly selected within each village, in a way that ensured two households included a pregnant or lactating mother or a married woman who was pregnant within the last two years, and the other two included children aged 6-15. The same households participated in the follow-up surveys which also used the original baseline questionnaire and respondent lists. The expansion of the programme post-2007 did not affect the composition of the control group to a large extent since new subdistricts, outside of the original sample, were prioritised for treatment. However, the value of the cash transfer fell from 14\% of monthly household consumption in 2007 to 7\% by 2013. 

\subsection{Related literature}
Existing evidence on the impacts of CCTs on health care utilisation is vast. Early impact evaluations of pioneering programmes implemented in Latin America and the Caribbean have generated substantial evidence on their effectiveness in increasing the utilisation of preventive health care services among the poor, and in some cases, improving health outcomes \citep{Lagarde2007-hb,Glassman2007-oe,Ranganathan2012-md}. For example, there were substantial increases in prenatal care visits of 8\% and 19\% in Mexico (\textit{Progresa}) and Honduras (\textit{Programe de Asignacion Familiar, PRAF} \citep{Barber2009-ii,Morris2004-cj}. Looking beyond average effects, \citet{Cooper2020-bx} conducted a review into the existing literature reporting heterogeneity in programme impacts across population subgroups defined according to sex, socioeconomic status, region and education. Of the 56 reviewed studies, 40 reported subgroup effects presented either as stratum-specific effects or as interactions between effect modifiers and the intervention. Using evidence from India (\textit{Janani Suraksha Yojana, JSY}) and Mexico (\textit{Oportunidades}), they found that positive programme effects on health care utilisation were generally larger among women that are younger (aged 15-24), more disadvantaged, less educated, rurally-based, and from regions where the CCT scheme was more rigorously implemented. 

In Indonesia, impact evaluations of PKH support these earlier findings that the cash incentives translate to greater health care demand. \citet{Cahyadi2020-lw} find dramatic short- and longer- term effects of PKH on various behaviours (even after correcting for multiple hypothesis testing): an increase in the average number of postnatal visits (0.8) in 2009; and increases in the probability of having a facility delivery (17\%) and a delivery assisted by a doctor or midwife (23\%) in 2013. The authors, however, do not explore varying impacts across the population. \citet{Kusuma2016-tx} similarly find encouraging effects on utilisation in 2009, including an increase in the proportion of women who had $\geq4$ prenatal visits (4\%), $\geq$2 postnatal visits (5\%), and a facility delivery (7\%). They explore whether utilisation effects vary for pregnant women that are indicated as high-risk, finding that proportions of prenatal visits and facility delivery decrease as risk increases. \citet{Alatas2011-di} also finds  substantial increases in the likelihood of beneficiary mothers completing $\geq4$ prenatal visits (13\%) and $\geq2$ postnatal visits (21\%). They additionally reported subgroup effects, finding that PKH effects on newborn-related health care utilisation are greater among urban and non-agriculturally based households where health care facilities are more accessible and available, and among female-led households. Unlike previous heterogeneity evidence which tends to consistently report greater effects mostly among the poorer population, this report finds larger increases in facility delivery and post-natal visits for better-off households. Finally, they report that mothers with some formal education are more likely to have a facility delivery and make post-natal visits, whereas those with no education are more likely to have an assisted delivery.

\subsection{Data}
\begin{figure}[!h]
\caption{Distribution of pre- and post-natal health visits, by enrolment status and year}
\label{fig:visits} 
\centering
\scalebox{0.8}{\input{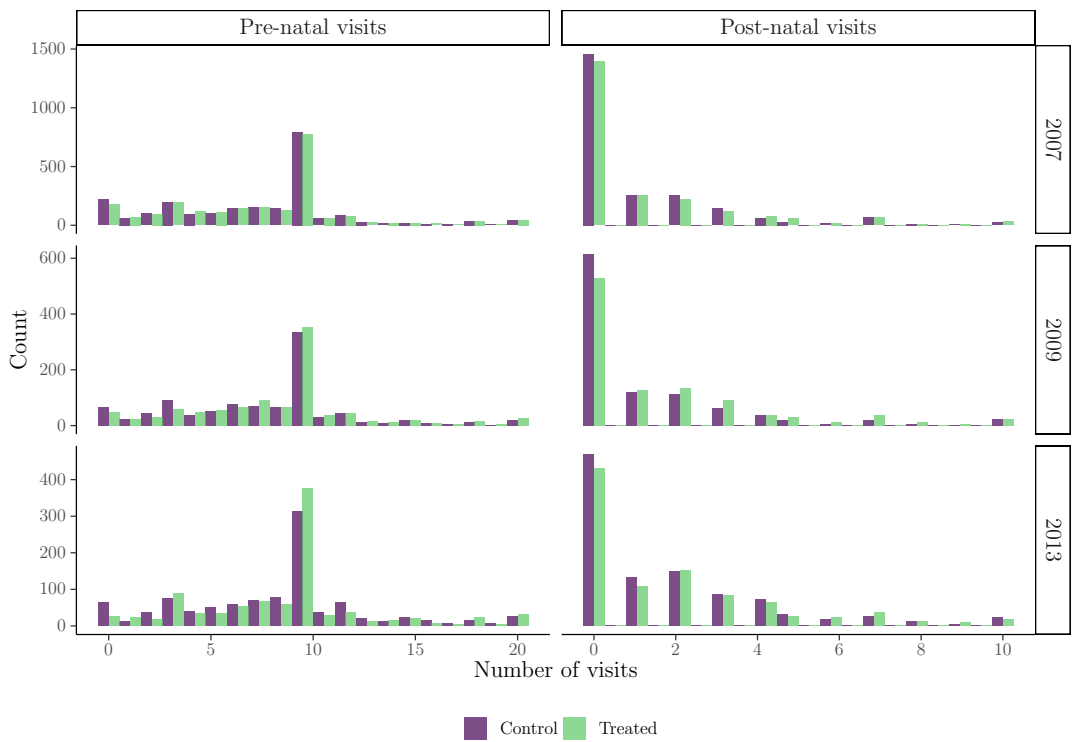}}
\begin{minipage}{0.9\linewidth} 
{\footnotesize \textit{Note: The main analysis in this paper covers survey waves from 2009 and 2013, but here we present pre-experiment outcomes from 2007 as a basis of comparison.}}
\end{minipage}
\end{figure}

We construct a dataset of married women aged 16 to 49 who had pregnancies or deliveries within the two years prior to the 2009 and 2013 follow-up surveys. For the outcomes, we construct four binary variables related to health care utilisation that indicate whether the woman attended at least four pre-natal check ups, the delivery took place at a medical facility, the delivery was assisted by a trained professional, and the woman attended at least two post-natal check ups. We decide to discretise the continuous outcomes for the number of pre- and post-natal visits since PKH requires a specified minimum number of visits to be met in order to make the cash transfer. Figure \ref{fig:visits} displays the distribution of the number of health visits made by control and treated populations both pre- and post-experiment. 

The surveys accompanying the randomised experiment collected individual, household and community-level information on demographic and socioeconomic characteristics (e.g. age, schooling and employment status), attributes of the household (e.g. household size, lack of utilities such as electricity and clean water), the local supply of health care services (e.g. number of practicing doctors), and information from village leaders' concerns related to healthcare provision.

These variables will be included in the estimation process to help adjust for observed confounding, since they explain eligibility and enrolment into PKH while also being independently associated with the outcomes.  We choose to include all of these variables as potential effect modifiers in our heterogeneity analyses, informed by results from past previous analyses, which find that the effects of PKH, and CCTs more generally, may vary across a set demographic, socioeconomic, geographic and supply-side variables \citep{Alatas2011-di}.  We remove observations with incomplete data on the outcomes and the vector of covariates. See Table \ref{table:sumstats} for the final sample sizes and unadjusted outcome means by enrolment status and year, and Table \ref{table:compliers} for the proportion of observations that comply with the randomisation protocol. In 2009, around half of those randomised to be in the programme were actually enrolled, while 10\% of those who were randomised to be in the control group were enrolled, and this increased to 14\% by 2013.

\begin{table}
\caption{Contingency table showing the association between the randomisation protocol and enrolment status}
\label{table:compliers}
\begin{minipage}[c]{0.5\textwidth}
\centering
\caption*{2009}
\resizebox{0.9\linewidth}{!}{
\begin{tabular}{>{\raggedright\arraybackslash}p{3cm}>{\centering\arraybackslash}p{2cm}>{\centering\arraybackslash}p{2cm}}
\toprule
\multicolumn{1}{c}{\textbf{}} & \multicolumn{2}{c}{\textbf{Randomisation ($Z$)}} \\
\cmidrule(l{3pt}r{3pt}){2-3}
  & Control & Treated\\
\midrule
\addlinespace[0.3em]
\multicolumn{3}{l}{\textbf{Enrolment ($D$)}}\\
\hspace{1em}Not enrolled & 929 (90\%) & 525 (51\%)\\
\hspace{1em}Enrolled & 99 (10\%) & 512 (49\%)\\
\bottomrule
\end{tabular}}
\end{minipage}
\begin{minipage}[c]{0.5\textwidth}
\centering
\caption*{2013}
\resizebox{0.9\linewidth}{!}{
\begin{tabular}{>{\raggedright\arraybackslash}p{3cm}>{\centering\arraybackslash}p{2cm}>{\centering\arraybackslash}p{2cm}}
\toprule
\multicolumn{1}{c}{} & \multicolumn{2}{c}{\textbf{Randomisation ($Z$)}} \\
\cmidrule(l{3pt}r{3pt}){2-3}
  & Control & Treated\\
\midrule
\addlinespace[0.3em]
\multicolumn{3}{l}{\textbf{Enrolment ($D$)}}\\
\hspace{1em}Not enrolled & 884 (86\%) & 501 (52\%)\\
\hspace{1em}Enrolled & 140 (14\%) & 464 (48\%)\\
\bottomrule
\end{tabular}}
\end{minipage}
\begin{tablenotes}
\footnotesize
\item Note: Randomisation protocol ($Z$) indicates whether the mother lives in a treatment or control subdistrict. Enrolment status indicates whether the mother actually received PKH (``enrolled") or not (``not enrolled").
\end{tablenotes}
\end{table}

\section{Methods}

\subsection{Estimation of Treatment Effects}

We are interested in separately estimating the causal effects of being enrolled into the PKH programme (compared to not being enrolled) in 2009 and 2013 on various outcomes relating to maternal health care utilisation - the number of pre-natal visits, the number of post-natal visits, the probability of an assisted delivery by a skilled midwife or doctor, and the probability of delivery at a health facility. We perform these analyses separately, using a common notation $Y$ for all outcomes, and $D$ for the binary indicator of PKH enrolment. Under the potential outcomes framework for causal inference, we denote the potential outcome that would be observed if individual $i$ was enrolled into programme $d$ by $Y_i(d)$. We define our two estimands of interest: (1) the average treatment effect (ATE), which takes the expectation of the individual treatment effects across the population, $\tau=E[Y_i(1)-Y_i(0)]$; and (2) the conditional average treatment effect (CATE), which evaluates the ATE for individuals with the same covariate profile $X_i=x$, $\tau(x)=E[Y_i(1)-Y_i(0)|X_i=x]$.

Following \citet{Athey2019-qf}, we define the relationship between $Y_i$ and $D_i$ using a structural model, $Y_i=m(X_i)+\tau(X_i)D_i+\varepsilon_i$, where $m(X_i)$ is a nuisance function whose shape is unspecified, and $\varepsilon_i$ is an error term. Since PKH was targeted to households (and not randomly assigned) and there was some reported non-compliance, we cannot proceed with the assumption that $\varepsilon_i$ is independent of $D_i$, meaning that a regression of $Y_i$ on $D_i$ will not yield a consistent estimate of $\tau(x)$. We introduce an instrumental variable $Z_i$, which is a binary indicator for whether the household is located in an initial PKH subdistrict. In other words, the IV represents the study randomisation mechanism. If $Z_i$ has a causal effect on $D_i$ conditionally on $X_i=x$ (the ``relevance" assumption), and affects $Y_i$ only through $D_i$ conditionally on $X_i$ (the ``exclusion restriction"), then $\tau(x)$ can be identified as follows:
\begin{equation}
\label{eq:1}
    \tau(x) = 
\frac{\mathrm{Cov}[Y,Z|X_i=x]}{\mathrm{Cov}[D,Z|X_i=x]},
\end{equation}
where the numerator is the conditional average intention-to-treat effect, interpreted as the conditional effect of being given the opportunity to enrol into PKH, and the denominator is the share of compliers in the sample (that is, the proportion of individuals that complied with the randomisation protocol). We can use heterogeneous treatment effect estimation methods that use the identification in (\ref{eq:1}) to estimate $\tau(x)$ as the conditional local average treatment effect (CLATE), by solving an estimation equation of the form:
\begin{equation}
\label{eq:2}
    E[\psi_{\tau(x),m(x)}(Y_i,D_i,Z_i)|X_i=x]=0 \;\textrm{for all} \; x\in \mathcal{X},
\end{equation}
where,
\begin{equation}
\label{eq:3}
    \psi_{\tau(x),m(x)}= \left(\begin{array}{lr}
        Z_i(Y_i-D_i\tau(x)-m(x))\\
        Y_i-D_i\tau(x)-m(x)\\
        \end{array}\right).
\end{equation}

In (\ref{eq:3}), the first row implies that the correlation between the instrument and the error term is zero, and the second row implies that the error term is mean zero. We choose to estimate $\tau(x)$ using an instrumental forest that estimates causal effects using conditional two-stage least squares (2SLS). The instrumental forest estimator relies on a generalised random forest framework to find small neighbourhoods of observations (leaves of a tree) where $\tau(x)$ is similar, by performing a 2SLS regression using the residualised versions of the outcomes $Y_i-m(X_i)$, the treatment assignment $D_i-e(X_i)$ (where $e(x)=P[D|X_i]$ is the treatment propensity), and the instrument $Z_i-g(X_i)$ (where $g(x)=P[Z|X_i]$ is the instrument propensity)\footnote{Following \citet{Robinson1988-gu}, residualisation helps to mimimise confounding bias due to observed covariates by partialling out the effects of $X_i$.}\footnote{Note that $m(x)$, $e(x)$ and $g(x)$ are collectively referred to as the ``nuisance parameters" since they are not primarily of interest but are required to estimate the target causal parameter. In an instrumental forest, the nuisance parameters are internally estimated using separate regression forests, a predictive machine learning algorithm.}. So-called instrumental (causal) trees \citep{Athey2019-qf} are formed by recursively partitioning the data into leaves in a way that maximises the within-leaf heterogeneity in treatment effects. The trees are formed using a sample splitting technique referred to as ``honesty", to avoid overfitting  \citep{Athey2019-qf}.  This procedure is repeated across many bootstrapped samples to limit noise arising from individual trees, with the tree ensemble representing the instrumental forest. Each observation $i$ is assigned a weight $\alpha_i(x)$ that is calculated based on the frequency with which $i$ is used to estimate $\tau$ at $x$, averaged across trees. The treatment effect estimator generates CLATEs $\hat{\tau}(x)$ by solving the estimating equation in (\ref{eq:2}) with weights from the instrumental forests.

Individual treatment effects $\hat{\tau}(X_i)$ are estimated by evaluating $\hat{\tau}(x)$ at each covariate profile $X_i$. The treatment effects $\hat{\tau}(X_i)$ can also be aggregated over the entire population to provide an estimate of the local average treatment effect (LATE), by plugging in $\hat{\tau}(X_i)$ into a variant of the augmented inverse probability of treatment weighted estimator (also known as the doubly robust estimator), formed by taking the average of so-called doubly robust scores $\Gamma_i$: 
\begin{equation}
    \hat{\tau} = \frac{1}{N}\sum^N_{i=1}\hat{\Gamma}_i,\quad \hat{\Gamma}_i =
   \hat{\tau}(X_i)
     +
     \frac{\left(\frac{Z_i-\hat{g}(X_i)}{\hat{g}(X_i)(1-\hat{g}(X_i))}\right)}{\delta(X_i)}(Y_i-\hat{m}(X_i)-(D_i-\hat{e}(X_i)\hat{\tau}(X_i)).
\end{equation}
Construction of this particular doubly robust score requires estimates of $m(x)$, $e(x)$ and $g(x)$, which are separately estimated via regression forests. It also requires a so-called compliance score $\delta(X)=E[D|X,Z=1]-E[D|X,Z=0]$, which is an estimate of the causal effect of $Z$ on $D$ that is estimated via an auxiliary causal forest.\footnote{Causal forests also rely on the generalised random forest framework but find neighbourhoods of observations where the CATEs are similar. Note that when the instrument $Z$ and treatment $D$ are the same, an instrumental forest is equivalent to a causal forest.}

\subsection{Inference on Treatment Effect Heterogeneity}

Once we have obtained estimates of the individual CLATEs and double robust scores for each individual, we can use these estimates to examine drivers of treatment effect heterogeneity in a data-driven way. One way to assess treatment effect heterogeneity is to perform a linear regression of the doubly robust scores $\Gamma_i$ on $X$ to compare the relative contribution of covariates in predicting the CLATEs \citep{Semenova2021-lp,Chernozhukov2018-rt}. The resulting coefficients from the linear model are referred to as the best linear predictors (BLP) of CLATEs. They have a \textit{ceteris paribus} interpretation but should not be interpreted as partial effects since the true $\tau(x)$ may not be linear in $X$. If a coefficient of the BLP for $X_i$ is positive and significant, we interpret as $X_i$ having a significant positive linear impact on the treatment effect heterogeneity, holding all other variables constant. 

Another way that Chernozhukov et al (2018) suggest assessing treatment effect heterogeneity is through use of a ``Classification Analysis" (CLANs). This involves partitioning data into quartiles according to the estimated double-robust scores $\Gamma_i$, in effect  ranking the observations from low to high estimated treatment effects. For each effect modifier of interest, we regress the variable on the indicator of being in the most affected group, using ordinary least squares (OLS). This analyses is then repeated for the indicator of being in the least affected group. For each effect modifier,  we test whether the difference between the two estimated coefficients is statistically significant.  If the difference is significantly positive, then those individuals with characteristic $X_i$ experienced greater levels of the treatment effect. In contrast to the BLP analysis, this can be interpreted as a univariate analysis - as effect modifiers are analysed one by one, without controlling for the others - and can provide further evidence of treatment effect heterogeneity with respect to specific covariates expressed as binary indicators.

Finally, we turn to a method commonly used to learn optimal treatment allocation rules in an interpretable way: policy trees. The policy tree learning algorithm performs exhaustive search over all possible trees using the estimated $\Gamma_i$, choosing as the final treatment rule the tree which maximises the overall treatment effect \citep{Athey2021-uo}. In other words, we estimate the optimal policy $\hat{\pi}(X_i)$ which maximises a value function $\hat{A_n}(\pi)$:

\begin{equation}
    \hat{\pi}_n = \argmax \left\{ \hat{A}_n(\pi) : \pi \in \Pi_n\right\},\quad 
    \hat{A}_n = \frac{1}{N}\sum^N_{i=1}(2\pi(X_i)-1)\hat{\Gamma}_i,
\end{equation}

where $\Pi$ is the class of binary decision rules. We are interested in whether those characteristics that show significant relationship to treatment effect heterogeneity using BLPs and CLANs are also most commonly used by the policy tree algorithm to assign treatment under the optimal policy $\hat{\pi}$. We report which variables are chosen as the most important decision criteria when assigning the optimal treatment regime. 

The analytical steps taken in this paper are described as follows:
\begin{enumerate}
    \item Train an instrumental forest using default settings (i.e. 2000 trees in the ensemble, and 200 trees to select tuning parameters) :
    \begin{itemize}
        \item Nuisance parameters -- $m(x)$, $e(x)$ and $g(x)$ -- are estimated using separate regression forests, where the propensity score $e(x)$ is estimated without including supply side variables in $X$. 
        \item The entire covariate vector $X$ is used for the recursive partitioning - see Table \ref{table:covariates} for a list of covariates.
    \end{itemize}

\begin{table}[!h]
\caption{List of selected variables in $X$}
\label{table:covariates}
\centering
\resizebox{\linewidth}{!}{
\begin{threeparttable}
\begin{tabular}[t]{l c}
\toprule
 Variable & Used in $\hat{m}(x)$\\
\midrule
Enrolled into subsidised insurance & Yes\\
Enrolled into other (non subsidised) insurance & Yes\\
Lives in Java & Yes \\
Lives in an urban area & Yes\\
Age (16-29; 30-39; 40-49) & Yes\\
Mother is educated (at elementary level) & Yes\\
Mother is employed & Yes\\
Head of household is educated (at elementary level) & Yes\\
Head of household is employed in agriculture sector & Yes\\
Head of household is employed in service sector & Yes\\
Household spends above average on alcohol and tobacco & Yes\\
Number of practising doctors in village per capita (q1-q3) & No\\
Number of practising nurses in village per capita (q1-q3) & No\\
Number of practising midwives in village per capita (q1-q3)& No\\
Number of practising traditional birthing attendants in village per capita (q1-q3) & No\\
Ln(size of household) (q1-q3) & Yes\\
Log(household non food expenditure per capita) (q1-q3) & Yes\\
Household has no clean water & Yes \\
Household has no own latrine & Yes \\
Household has no septic tank & Yes \\
Household has no electricity & Yes\\
Village chief indicates ``lack of healthcare facilities" a top 3 concern & No \\
Village chief indicates ``lack of medical equipment" a top 3 concern & No\\
Village chief indicates ``low healthcare awareness" a top 3 concern & No \\
\bottomrule
\end{tabular}
\begin{tablenotes}
\item \textit{Note: Continuous variables have been discretised into terciles (q1-q3).} 
\end{tablenotes}
\end{threeparttable}}
\end{table}
    \item Predict $\hat{\tau}(X_i)$ by evaluating the trained instrumental forest for each observation's covariate profile:
    \begin{itemize}
        \item Predictions are made ``out-of-bag", meaning that only the trees that did not use observation $i$ during the training process are used in the prediction.\footnote{Out-of-bag prediction produces CLATE estimates without the need for an additional data splitting step \citep{Athey2019-qf}. Standard errors of the predicted CLATEs are clustered at the subdistrict level; the level of randomisation.}
    \end{itemize}
    \item Construct doubly robust scores $\hat{\Gamma}_i$ to compute the LATE $\tau$.
    \item Assess the treatment effect heterogeneity captured by the forest outputs:
    \begin{itemize}
        \item Plot a histogram of the estimated CLATEs $\hat{\tau}(X_i)$.
        \item Perform a linear regression of $\hat{\Gamma}_i$ on $X_i$ to find the best linear predictors of the CLATEs, and plot the estimated regression coefficients. 
        \item Test the difference between the most and least effected individuals for each variable (CLANs), and plot the estimated differences, with confidence intervals. 
        \item Learn and plot a depth-two policy tree to examine which covariates are chosen as most important splitting criteria.  
    \end{itemize}
\end{enumerate}

We implement these steps for each outcome and year separately.

\section{Results}

Table \ref{table:sumstats} presents summary statistics for our sample populations in 2009 and 2013. We report covariate means according to the the randomisation status $Z$ (we refer to these in treated and control samples), defined as those living in subdistricts that are assigned versus not assigned to PKH. We also report covariate means for sample populations based on their actual enrollment status in the PKH, $D$ (referring to these as enrolled and not enrolled samples). Overall, the table highlights that the target population is largely rural-based (approximately 90\%) with household heads who work in agriculture, and that the majority (50-60\%) live on the island of Java and lack household utilities such as running water. 

As expected, the average characteristics for treated and control samples are similar given the random assignment of PKH to subdistricts, with none of the reported SMDs being greater than 0.1. When comparing enrolled and not enrolled mothers, we find some large differences, painting of pictures of enrolled mothers being typically of worth socioeconomic status. For example, in both time periods, enrolled mothers are less likely to live in Java, have larger households, spend more on non-food items, and have a greater supply of traditional birth attendants in the village, compared to non-enrolled mothers. Between 2009 and 2013, we can observe an an increase in non-compliance (only 77\% of mothers living in subdistricts that are assigned to PKH were actually enrolled in 2013, compared to 84\% in 2009 (see Table \ref{table:compliers}), leading to further increase in the imbalance between enrolled and not enrolled mothers. This increased imbalance is most notable for supply-side variables,  somewhat reducing the relative disadvantage of the enrolled group: we find that enrolled women live in villages that are  more likely to have a greater supply of doctors and nurses per capita, than those not enrolled. 

We can also use this descriptive information to contrast the characteristics of the compliers to the randomised population. The compliers - those who actually enrol into PKH - are more likely to be older (aged 30-49) and have subsidised health insurance. They are also  more likely to be urban-based but less likely to live in Java, and they tend to have slightly larger households compared to the randomised population.

\subsection{Local Average Treatment Effects}

Figure \ref{fig:histogram} uses histograms to show the distribution of the estimated CLATEs ($\hat{\tau}(x)$ from the instrumental forest for each outcome and year).\footnote{Point estimates and standard errors are reported in the Appendix.} Looking firstly at local average effects, the programme has positive significant short- and longer-term impacts on the probability that the mother has a good assisted delivery (LATE=0.15 (SE=0.07) in 2009, LATE=0.16 (SE=0.07) in 2013). The beneficial average impacts on the probability that the mother meets the required threshold for pre- and post-natal visits are only significant in 2009 (pre-natal LATE=0.16 (SE=0.06), post-natal LATE=0.22 (SE=0.07)), suggesting that the programme has more immediate rather than sustained effects on health care visits.\footnote{Appendix Figure \ref{fig:histogram.visits} presents the distribution of the estimated CLATEs for the continuous versions of the health visits outcomes, which finds a positive average effect on post-natal visits only (LATE=0.95 (SE=0.33)).} Lastly, we find no effect on the probability that the mother has a facility delivery in both time periods. Looking beyond average effects, the histograms provide evidence of treatment effect heterogeneity since for each outcome and year combination, the CLATEs are not just distributed around the LATEs but span both negative and positive values, suggesting that the CCT programme is successful in incentivising some but not all compliers to increase their demand for maternal health care. In particular, the impact on post-natal visits in 2013 is the most variable among all outcomes, with CLATEs ranging between -0.5 to 0.7, implying that some proportion of compliers are less likely to attend at least 2 post-natal check ups after receiving the cash transfer. On the other hand, the impact on the delivery taking place at a medical facility shows the least heterogeneity, with CLATEs ranging between -0.1 and 0.5, so in this case a smaller proportion of compliers display adverse behaviour in response to the programme. 

\subsection{Best Linear Predictors of Treatment Effects}

Figures \ref{fig:blp} plots the estimated coefficients from the best linear predictor analysis that linearly regresses the doubly robust scores $\hat{\Gamma}_i$ on $X_i$.\footnote{Figure \ref{fig:blp.visits} plots coefficients for the continuous outcome.} We report the results for each outcome in turn. Starting with the good assisted delivery outcome, the binary indicator for the head of household working in the agriculture sector is a negative predictor of treatment effects in 2009. For 2013, household lack of septic tank and electricity are associated with an increase in treatment effect, as is the village head naming the ``lack of medical equipment" as a primary concern. This suggests that the effectiveness of PKH on good assisted delivery outcomes in 2013 could be driven by those most poor households, and improving access to health facilites for those mothers located in areas with low supply. For the facility delivery we find that residing in an urban location (but not Java) is a strong positive predictor of an increase in facility delivery in 2009, as is the village chief indication of ``low health awareness" as a major concern. Negative predictors of the treatment effect were only present in 2013, and only included household lack of a latrine. However, lack of electricity was again a positive predictor, indicating that the PKH may have incentivised mothers who lacked appropriate provisions for a home birth to have medical facility delivery. 

For the pre-natal visits outcome, we find a negative coefficient on the indicator for not having a septic tank in 2009, as well as for being in the highest tercile of nurses per-capita, which could be explained by the fact that mothers already living in villages with a higher supply of nurses per capita, compared to villages with the lowest supply, may not change their health care demand in response to the CCT programme. For 2013, the household having no electicity is a positive predictor of treatment effects, and the ``lack of healthcare facilities" village concern variable is negatively associated treatment effects. Taken together, these results point towards a potential prohibitive role of  travel time for women living in villages with no appropriate health care facilities, preventing them from enjoying the incentives provided by the PKH. Finally, for post-natal visits, we find that being in the highest tercile of midwives had a negative effect in 2009. In 2013, being in the second tercile of nurses had a negative effect, but household lack of clean water and electricity had a positive effect.

\subsection{Classification Analysis}

We now turn our attention to interpreting the CLAN results. A positive sign for a given coefficient indicates that individuals in the group (e.g. urban residents) experience higher treatment effects compared to those not in the group (e.g. rural residents). Conversely, a negative sign indicates that those in the group have lower treatment effects relative to those outside the group. It is important to note that this analysis examines each effect modifier separately, without controlling for other variables, and focuses on the univariate relationship between group membership and treatment effect. For the assisted delivery outcome, we find that urban residents had a higher treatment effect compared to rural residents in 2009, but this difference was no longer observed in 2013. Those employed in the agriculture sector experience lower levels of the treatment effect in 2009 as well. Taken together, these findings indicate that the initial impacts of the programme may have been driven by those residing in urban locations (as agriculture workers are less likely to live in these areas). For all of the supply-side variables (doctors, nurses, and midwives per capita), we see that those living with the lowest tercile had the highest treatment effects in 2013. For the facility delivery outcome, this relationship is also true for those living in the lowest tercile of traditional birth attendants. The household residing in Java was associated with higher levels of the treatment effect, but lack of a household latrine, and low village medical awareness, were associated with lowest levels of the treatment effect for facility delivery in 2013. 

For the pre-natal visits outcome, we see major differences in the CLAN results between survey waves. In 2013, for example, many of the supply variables (highest terciles of nurses and midwives, lack of facilities and medical equipment) are associated with low levels of the treatment effect, in contrast to what we observe for 2009. We also see living in Java associated with a high levels of hetreatment effects for prenatal visits in 2013, indicating that long-term incentivising for these visits may be more successful in areas with adequate transportation infrastructure. For the post-natal visits outcome, we see that households without a septic tank have lower levels of the treatment effect, as do those employed in the agriculture sector. However, those in the agriculture sector experience significantly higher levels of the treatment effect in 2013, as do those living in the lowest terciles of nurses and doctors (but highest tercile of midwives). These results could be explained by differences in the type of personnel most likely to be present for a pre- or post-natal visit (doctor or nurse vs midwive).  

\subsection{Policy Trees}

\begin{landscape}
\begin{figure}[!h]
\caption{Depth-two policy trees learned from $\Gamma_i$.}
\label{fig:policytrees}
\centering
\scalebox{0.9}{\includegraphics{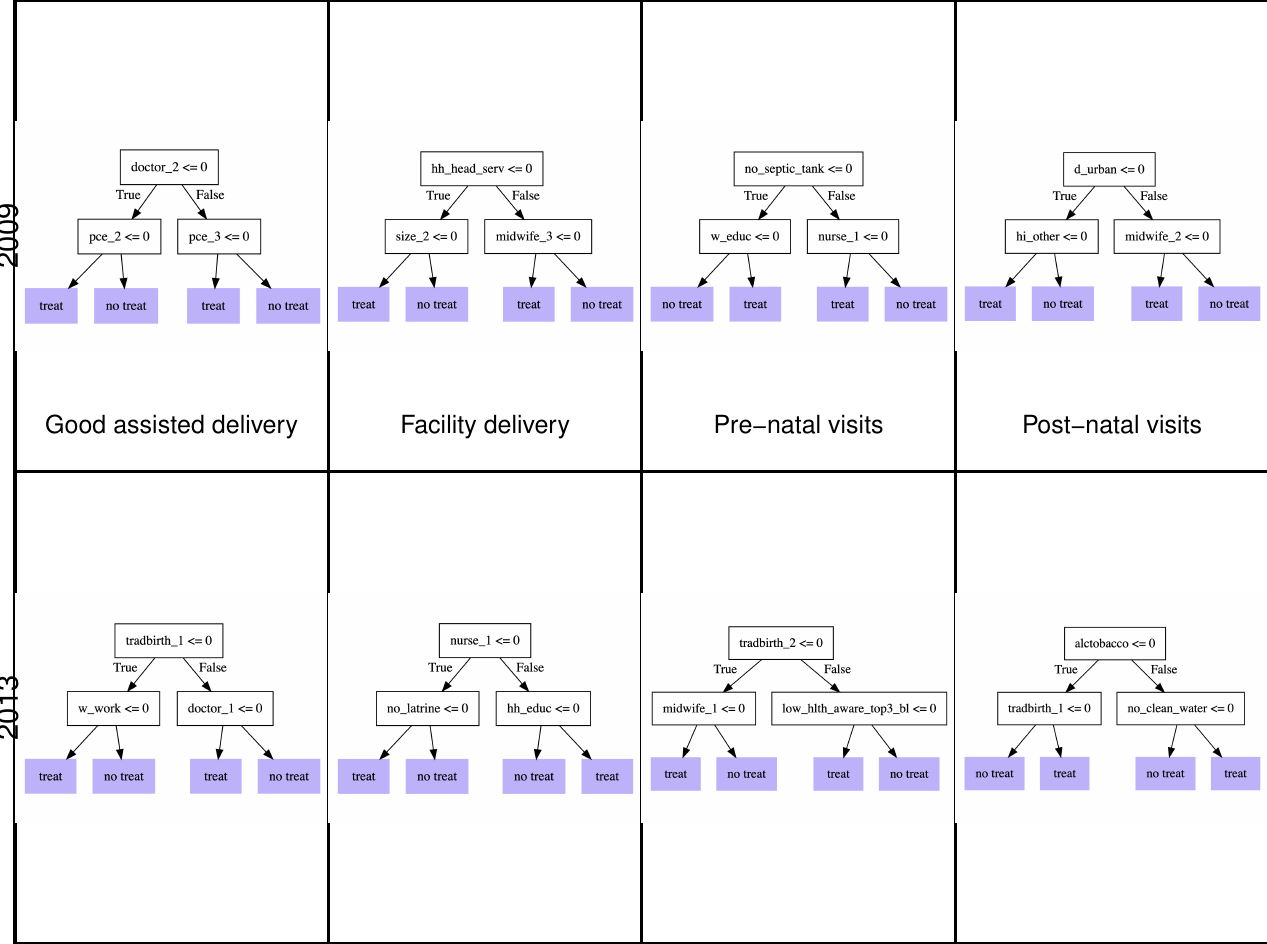}}
\begin{minipage}{0.9\linewidth} 
{\footnotesize \textit{Note: The top row presents trees for 2009, and the bottom row for 2013. Each column depicts one of the four outcomes.}}
\end{minipage}
\end{figure}%
\end{landscape}

We now examine the depth-2 decision trees learned from the estimated double robust scores $\hat{\Gamma}_i$. The trees are depicted in Figure \ref{fig:policytrees}, with each row representing a survey year and each column representing one of the four outcomes under analysis. In all eight decision trees, there is an importance of healthcare worker supply in terms of decision criteria. For 2013, the top decision node in three of the four outcomes is related to health worker supply. For the good assisted delivery outcome, for example, the first decision criteria is whether there are a high number of traditional birth attendants, and the right bottom node depends on whether there is a high supply of doctors. 

There are major differences in learned trees between the two survey waves. In 2009, the most important decision criteria for all four outcomes included mother's education and health insurance status, the household per-capita expenditure, size, urban location, and lack of septic tank, as well as supply of all types of healthcare personnel. In 2013, there is slightly less influence of the mother and household characteristics on the decision rules, with the nodes being for maternal employment, lack of latrine, electricity or clean water, and household head education and alcohol/tobacco expenditure. All four health worker supply variables were important splitting criteria for the optimal treatment allocation. 

Taken together and examined qualitatively, the policy tree results indicate a strong influence of the supply-side readiness of each village in terms of maximising desired maternal healthcare demand. The results also suggest that effects are important for those poorer households in urban locations and with fewer household amenities, indicating that the PKH programme was successful in incentivising those poorest participants. For example, in 2013 the bottom left decision node for the post-natal visits outcome assigns treatment to those who do not have clean water (no clean water $\leq 0$ is false $\rightarrow$ treat).

\section{Discussion}
In this paper, we used data on new mothers from a randomised experiment to evaluate the local average and heterogeneous effects of the PKH programme on various maternal health care utilisation outcomes in 2009 and 2013. We used a causal machine learning method, instrumental forests, to estimate heterogenous treatment effects (CLATEs), and aggregated these estimates over the entire sample population to produce a doubly robust approximation to the LATE. We also performed three types of complementary analysis of the drivers of heterogenous treatment effects: explored the best linear predictors of treatment effects, conducted a classification analysis, and built interpretable policy trees.

Our results largely support those from early evaluations on the overall average impacts of PKH on (compliant) new mothers, with increases in the probabilities of having a good assisted delivery in 2009 and 2013, attending at least four pre-natal check ups in 2009, and attending at least two post-natal check ups in 2009. However, the sizes of the effects tend to vary across studies, which can be explained by variations in study designs resulting in differences in covariate selection and identification of causal effects. Beyond average effects, the distribution of CLATEs provides evidence of heterogeneity in treatment effects such that although most mothers are expected to increase health care demand in response to the cash transfer based on their observed characteristics, others are less affected. 

Our analysis of drivers of treatment effect heterogeneity suggests that location and supply-side factors are important determinants of varying treatment effects for several outcomes. Urban-based households, where health care supply is more readily available, due to better proximity of medical facilities and a greater supply of practicing health care workers, are less likely to change their demand for maternal health care in response to the cash transfer. Other related variables, such as whether the household is located in Java and the nature of employment of the head-of-household, which is inherently linked to geographical factors, are also identified to be important predictors. We find that for the most part, the estimated regression coefficients from the BLP and the CLAN analysis are significant for one time period only, either 2009 or 2013, with only a few maintaining their significance throughout both periods, indicating a changing role of characteristics in programme effectiveness over time.

Our study may provide some insights into the factors affecting the duration and distribution of policy effects. The finding that PKH is unable to consistently maintain effectiveness beyond the short-term, if at all, could be explained by some reported issues in programme design and implementation \citep{Kusuma2016-tx}. Administrative problems resulting in payment delays and missed payments altogether could partly explain the limited impact, combined with the fact that cash payments (as a proportion of household consumption) were essentially halved between 2007 and 2013, thus significantly reducing the incentive-based component of the policy. Our results also suggest that geographical factors that are inherently linked to health care supply are important predictors of heterogeneity in policy impacts. Although PKH aims to target poor households in supply-ready areas, residual differences in health care accessibility and availability seem to contribute to varying policy impacts. It has been argued that, in addition to supply-side readiness, other contextual differences, including cultural factors and supply-side barriers, can impact programme effectiveness \citep{Glassman2013-xo}. For example, poor quality of care, transportation costs and a lack of health knowledge or programme awareness may restrict health care use irrespective of the value of the cash payment or the availability of health facilities in the local area \citep{Gaarder2010-ss}. These findings suggest the need to better align demand-side policies with supply-side initiatives to support policy effectiveness.

\clearpage
\begin{landscape}\begin{table}[!h]
\caption{Summary statistics}
\label{table:sumstats}
\centering
\begin{adjustbox}{width=0.8\linewidth}
\begin{threeparttable}
\begin{tabular}{>{\raggedright\arraybackslash}p{5cm}>{\centering\arraybackslash}p{2cm}>{\centering\arraybackslash}p{2cm}>{\centering\arraybackslash}p{2cm}>{\centering\arraybackslash}p{2cm}>{\centering\arraybackslash}p{2cm}>{\centering\arraybackslash}p{2cm}>{\centering\arraybackslash}p{2cm}>{\centering\arraybackslash}p{2cm}>{\centering\arraybackslash}p{2cm}>{\centering\arraybackslash}p{2cm}>{\centering\arraybackslash}p{2cm}>{\centering\arraybackslash}p{2cm}}
\toprule
\multicolumn{1}{c}{\textbf{}} & \multicolumn{6}{c}{\textbf{\makecell[c]{2009\\(n=2,065)}}} & \multicolumn{6}{c}{\textbf{\makecell[c]{2013\\(n=1,989)}}} \\
\cmidrule(l{3pt}r{3pt}){2-7} \cmidrule(l{3pt}r{3pt}){8-13}
\multicolumn{1}{c}{} & \multicolumn{3}{c}{Randomisation} & \multicolumn{3}{c}{Enrolment} & \multicolumn{3}{c}{Randomisation} & \multicolumn{3}{c}{Enrolment} \\
\cmidrule(l{3pt}r{3pt}){2-4} \cmidrule(l{3pt}r{3pt}){5-7} \cmidrule(l{3pt}r{3pt}){8-10} \cmidrule(l{3pt}r{3pt}){11-13}
  & Control & Treated & SMD & Not enrolled & Enrolled & SMD & Control & Treated & SMD & Not enrolled & Enrolled & SMD\\
\midrule
\addlinespace[0.3em]
\multicolumn{13}{l}{\textbf{Outcomes}}\\
\hspace{1em}Pre-natal visits $\geq4$ & 0.644 & 0.700 & 0.056 & 0.677 & 0.661 & -0.016 & 0.771 & 0.838 & 0.067 & 0.805 & 0.801 & -0.004\\
\hspace{1em}Good assisted delivery & 0.460 & 0.517 & 0.057 & 0.494 & 0.475 & -0.020 & 0.721 & 0.768 & 0.047 & 0.755 & 0.719 & -0.036\\
\hspace{1em}Facility delivery & 0.778 & 0.846 & 0.067 & 0.803 & 0.835 & 0.032 & 0.817 & 0.838 & 0.021 & 0.838 & 0.805 & -0.033\\
\hspace{1em}Post-natal visits$\geq2$ & 0.286 & 0.366 & 0.080 & 0.309 & 0.368 & 0.059 & 0.412 & 0.441 & 0.029 & 0.436 & 0.404 & -0.032\\
\addlinespace[0.3em]
\multicolumn{13}{l}{\textbf{Mother's characteristics}}\\
\hspace{1em}Age:16-29 & 0.408 & 0.467 & 0.059 & 0.457 & 0.390 & -0.068 & 0.537 & 0.578 & 0.041 & 0.570 & 0.528 & -0.042\\
\hspace{1em}Age:30-39 & 0.488 & 0.436 & -0.052 & 0.452 & 0.486 & 0.034 & 0.368 & 0.333 & -0.036 & 0.351 & 0.351 & 0.000\\
\hspace{1em}Age:40-49 & 0.104 & 0.097 & -0.007 & 0.091 & 0.124 & 0.034 & 0.095 & 0.089 & -0.006 & 0.079 & 0.121 & 0.041\\
\hspace{1em}Educated (elementary) & 0.802 & 0.828 & 0.027 & 0.818 & 0.809 & -0.009 & 0.861 & 0.876 & 0.014 & 0.879 & 0.843 & -0.037\\
\hspace{1em}Employed & 0.215 & 0.200 & -0.015 & 0.199 & 0.226 & 0.026 & 0.237 & 0.187 & -0.051 & 0.209 & 0.220 & 0.011\\
\hspace{1em}Subsidized insurance & 0.729 & 0.747 & 0.019 & 0.710 & 0.804 & 0.093 & 0.738 & 0.759 & 0.020 & 0.716 & 0.821 & 0.105\\
\hspace{1em}Other insurance & 0.046 & 0.054 & 0.008 & 0.053 & 0.043 & -0.010 & 0.050 & 0.062 & 0.012 & 0.059 & 0.048 & -0.011\\
\addlinespace[0.3em]
\multicolumn{13}{l}{\textbf{Head of household's characteristics}}\\
\hspace{1em}Educated(elementary) & 0.696 & 0.728 & 0.033 & 0.710 & 0.715 & 0.005 & 0.775 & 0.779 & 0.004 & 0.791 & 0.747 & -0.044\\
\hspace{1em}Works in agriculture  & 0.591 & 0.628 & 0.036 & 0.585 & 0.668 & 0.082 & 0.534 & 0.508 & -0.026 & 0.510 & 0.548 & 0.038\\
\hspace{1em}Works in service sector & 0.164 & 0.157 & -0.007 & 0.162 & 0.159 & -0.003 & 0.111 & 0.089 & -0.022 & 0.108 & 0.083 & -0.026\\
\addlinespace[0.3em]
\multicolumn{13}{l}{\textbf{Household characteristics}}\\
\hspace{1em}Lives in Java & 0.611 & 0.621 & 0.010 & 0.662 & 0.506 & -0.157 & 0.681 & 0.656 & -0.025 & 0.725 & 0.540 & -0.185\\
\hspace{1em}Urban location & 0.129 & 0.115 & -0.015 & 0.117 & 0.134 & 0.017 & 0.120 & 0.119 & -0.001 & 0.115 & 0.131 & 0.016\\
\hspace{1em}Num. HH members (ln) & 1.813 & 1.793 & -0.061 & 1.796 & 1.821 & 0.075 & 1.817 & 1.818 & 0.003 & 1.804 & 1.850 & 0.132\\
\hspace{1em}Non-food exp (PC) & 10.918 & 10.933 & 0.024 & 10.965 & 10.832 & -0.204 & 11.521 & 11.569 & 0.072 & 11.581 & 11.459 & -0.180\\
\hspace{1em}Alcohol/tobac exp (PC) & 7.638 & 7.871 & 0.062 & 7.741 & 7.790 & 0.013 & 8.386 & 8.705 & 0.082 & 8.387 & 8.892 & 0.135\\
\hspace{1em}No clean water & 0.869 & 0.891 & 0.022 & 0.881 & 0.877 & -0.004 & 0.892 & 0.875 & -0.017 & 0.886 & 0.877 & -0.008\\
\hspace{1em}No latrine & 0.483 & 0.505 & 0.022 & 0.460 & 0.576 & 0.116 & 0.456 & 0.452 & -0.004 & 0.418 & 0.536 & 0.118\\
\hspace{1em}No septic tank & 0.655 & 0.661 & 0.006 & 0.639 & 0.702 & 0.063 & 0.582 & 0.590 & 0.008 & 0.565 & 0.634 & 0.069\\
\hspace{1em}No electricity & 0.226 & 0.202 & -0.024 & 0.184 & 0.283 & 0.099 & 0.112 & 0.115 & 0.003 & 0.090 & 0.167 & 0.077\\
\addlinespace[0.3em]
\multicolumn{13}{l}{\textbf{Health care supply (number of practising workers in village per capita*1000)}}\\
\hspace{1em}Doctors & 0.321 & 0.301 & -0.025 & 0.301 & 0.334 & 0.043 & 0.266 & 0.259 & -0.011 & 0.239 & 0.318 & 0.123\\
\hspace{1em}Nurses & 0.591 & 0.585 & -0.005 & 0.564 & 0.644 & 0.058 & 0.516 & 0.519 & 0.002 & 0.484 & 0.593 & 0.106\\
\hspace{1em}Midwives & 0.523 & 0.518 & -0.007 & 0.507 & 0.553 & 0.068 & 0.496 & 0.510 & 0.025 & 0.494 & 0.524 & 0.053\\
\hspace{1em}Trad. birth attendants & 1.030 & 1.061 & 0.019 & 0.935 & 1.309 & 0.209 & 0.965 & 1.034 & 0.044 & 0.890 & 1.247 & 0.217\\
\addlinespace[0.3em]
\multicolumn{13}{l}{\textbf{Village characteristics (village head identified as top 3 concern)}}\\
\hspace{1em}Lack healthcare facilities & 0.312 & 0.299 & -0.013 & 0.309 & 0.298 & -0.011 & 0.271 & 0.293 & 0.022 & 0.266 & 0.318 & 0.051\\
\hspace{1em}Lack medical equipment & 0.170 & 0.211 & 0.041 & 0.175 & 0.229 & 0.054 & 0.159 & 0.193 & 0.034 & 0.170 & 0.189 & 0.019\\
\hspace{1em}Low health awareness & 0.132 & 0.095 & -0.037 & 0.115 & 0.111 & -0.004 & 0.110 & 0.101 & -0.010 & 0.103 & 0.113 & 0.010\\
\bottomrule
\end{tabular}\begin{tablenotes}
\item \textit{Note: All columns apart from SMD report covariate means. SMD = standardised mean difference. q1 = highest quantity. Observations with complete data included only.} 
\end{tablenotes}
\end{threeparttable}
\end{adjustbox}
\end{table}
\end{landscape}

\begin{landscape}
\begin{figure}[!h]
\caption{Histograms of estimated CLATEs $\hat{\tau}(x)$ from the instrumental forest, by outcome and year}
\label{fig:histogram} 
\centering
\scalebox{0.95}{\input{Paper/figs/hist.tex}}
\begin{minipage}{0.9\linewidth} 
{\footnotesize \textit{Note: Dashed lines denote the ATE point estimates and 95\% confidence intervals (via the AIPTW estimator). Red solid line at zero.}}
\end{minipage}
\end{figure}
\end{landscape}

\begin{landscape}
\begin{figure}[!h]
\caption{Estimated coefficients (and 95\% CIs) from the best linear predictor analysis of $\Gamma_i$ on $X_i$.}
\label{fig:blp}
\centering
\scalebox{0.8}{\input{Paper/figs/blp}}
\begin{minipage}{0.9\linewidth} 
{\footnotesize \textit{Note: Instrumental forest estimate of $\tau(x)$ using the instrument $Z_i$ as treatment $D_i$. HoH = head-of-household. HH = household. PCE = per capita expenditure. Tradbirth = traditional birth attendant. Continuous variables have been converted to discrete variables using terciles. Reference categories include: Age 40-49, Doctor:q3 (third tercile), Nurse:q3, Midwife:q3, Tradbirth:q3, HHsize:q3 and PCE:q3. q1 = largest quantity.}}
\end{minipage}
\end{figure}%
\end{landscape}

\begin{landscape}
\begin{figure}[!h]
\caption{Mean differences (and 95\% CIs) from the classification analysis (CLAN) of $\Gamma_i$.}
\label{fig:clan}
\centering
\scalebox{0.8}{\input{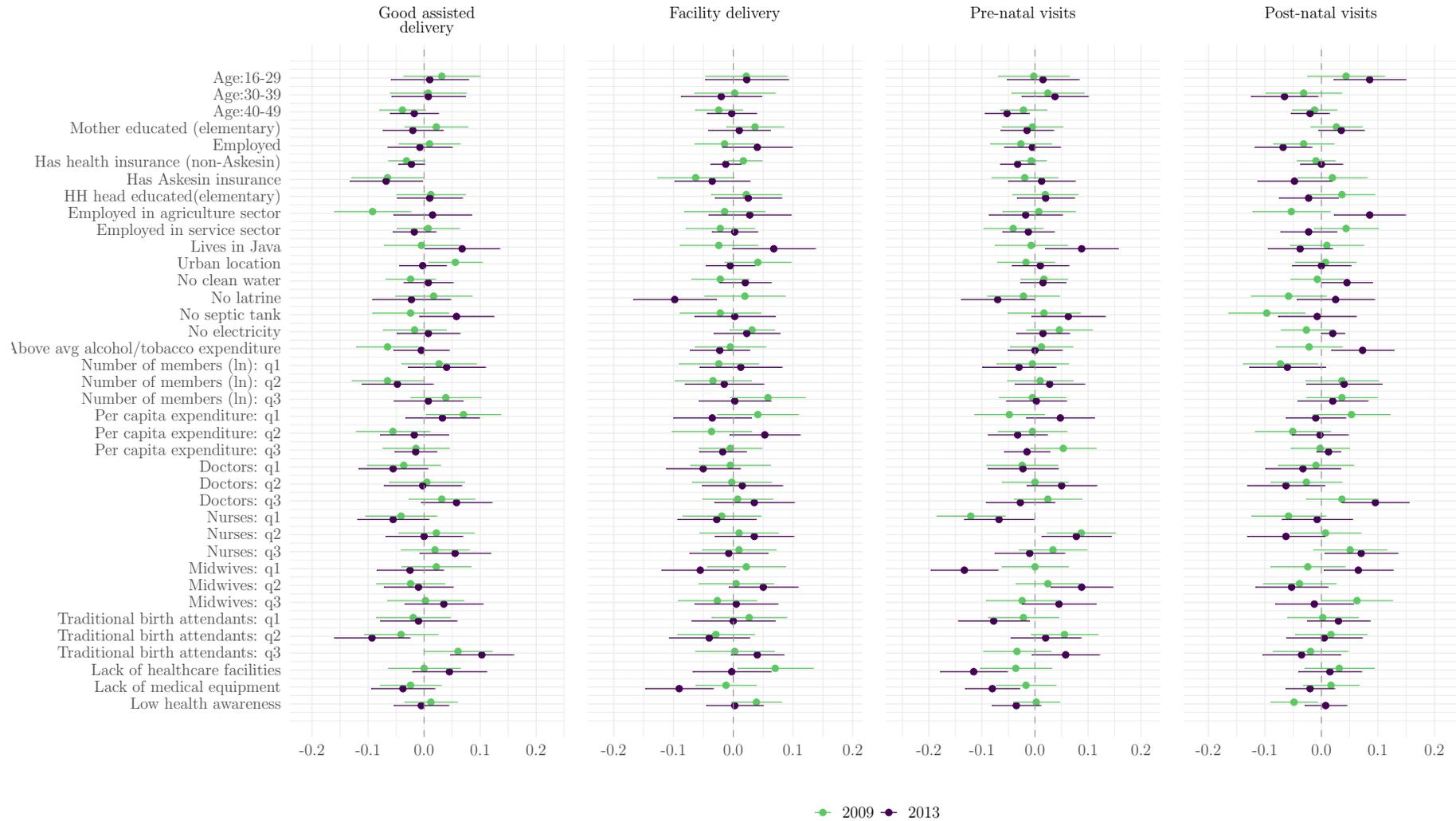}}
\begin{minipage}{0.9\linewidth} 
{\footnotesize \textit{Note: Effect modifiers are regressed on indicators of being in the high or low treatment effect groups. HoH = head-of-household. HH = household. PCE = per capita expenditure. Tradbirth = traditional birth attendant. Continuous variables have been converted to discrete variables using terciles. Note this is a univariate analysis, and thus there are no reference categories.}}
\end{minipage}
\end{figure}%
\end{landscape}

\clearpage 
\bibliography{Bibliography/paperpile}
\newpage
\appendix
\renewcommand\thefigure{\thesection.\arabic{figure}}
\renewcommand\thetable{\thesection.\arabic{table}}

\section{Estimates}

\begin{table}[ht]
\caption{ATEs and standard errors, all outcomes}
\centering
\setlength{\tabcolsep}{10pt} 
\begin{tabular}{l |rrr}
  \hline Year & Outcome & ATE & se \\ 
  \hline
2009 & Good assisted
delivery & 0.15 & 0.07 \\ 
2009 & Facility delivery & 0.15 & 0.09 \\ 
2009 & Pre-natal visits & 0.16 & 0.06 \\ 
2009 & Post-natal visits & 0.22 & 0.07 \\ 
\hline 
2013 & Good assisted
delivery & 0.16 & 0.07 \\ 
2013 & Facility delivery & 0.12 & 0.09 \\ 
2013 & Pre-natal visits & 0.07 & 0.07 \\ 
2013 & Post-natal visits & 0.09 & 0.08 \\ 
   \hline
\end{tabular}
\end{table}

\section{Robustness checks}
\subsection{Continuous version of pre/post-natal visits outcomes}
\begin{figure}[!h]
\caption{Histograms of estimated CLATEs $\hat{\tau}(x)$ from the instrumental forest, by outcome and year}
\label{fig:histogram.visits} 
\centering
\scalebox{0.9}{\input{Paper/figs/hist.visits}}
\begin{minipage}{0.9\linewidth} 
{\footnotesize \textit{Note: Dashed lines denote the ATE point estimates and 95\% confidence intervals (via the AIPTW estimator). Red solid line at zero.}}
\end{minipage}
\end{figure}

\begin{landscape}
\begin{figure}[!h]
\caption{Estimated coefficients (and 95\% CIs) from the best linear predictor analysis of $\Gamma_i$ on $X_i$.}
\label{fig:blp.visits}
\centering
\scalebox{0.8}{\input{Paper/figs/blp.visits}}
\begin{minipage}{0.9\linewidth} 
{\footnotesize \textit{Note: Instrumental forest estimate of $\tau(x)$ using the instrument $Z_i$ as treatment $D_i$. HoH = head-of-household. HH = household. PCE = per capita expenditure. Tradbirth = traditional birth attendant. Continuous variables have been converted to discrete variables using quintiles. Reference categories include: Age 40-49, Doctor:q4, Nurse:q4, Midwife:q4, Tradbirth:q4, HHsize:q4 and PCE:q4. q1 = largest quantity.}}
\end{minipage}
\end{figure}%
\end{landscape}
\end{document}